\documentclass[10pt,letterpaper,twocolumn]{article}

\usepackage{usenix,epsfig,endnotes}
\usepackage[square,comma,numbers,sort&compress]{natbib}

\usepackage{silence}
\PassOptionsToPackage{hyphens}{url}
\usepackage[hyphens]{url}
\usepackage[breaklinks,colorlinks]{hyperref}
\usepackage[usenames,dvipsnames]{xcolor}
\hypersetup{citecolor=blue,linkcolor=blue}
\usepackage{amsmath,amsopn,amssymb,mathtools}
\usepackage{endnotes,microtype,xspace,graphicx,fancyvrb,multirow}
\usepackage{booktabs}
\usepackage{array,underscore,relsize}
\usepackage[T1]{fontenc}
\usepackage{times}
\usepackage{fancyhdr,lastpage}
\usepackage{enumitem}
\usepackage[linesnumbered,algoruled,boxed,lined]{algorithm2e}
\usepackage[labelfont=bf,font=small,skip=5pt]{caption}
\usepackage{xfrac}
\usepackage{balance}

\usepackage{titling}

\usepackage[compact,small]{titlesec}
\titlespacing*{\section}{0pt}{0.2\baselineskip}{0.2\baselineskip}
\titlespacing*{\subsection}{0pt}{0.15\baselineskip}{0.15\baselineskip}

\setlist[itemize]{itemsep=1pt, topsep=2pt}

\newcommand{\sys}{\mbox{\textsc{Bunshin}}\xspace}
\newcommand{\srcdist}{\textit{check distribution}\xspace}
\newcommand{\secdist}{\textit{sanitizer distribution}\xspace}
\newcommand{\Srcdist}{\textit{Check distribution}\xspace}
\newcommand{\Secdist}{\textit{Sanitizer distribution}\xspace}
\newcommand{\SrcDist}{\textit{Check Distribution}\xspace}
\newcommand{\SecDist}{\textit{Sanitizer Distribution}\xspace}

\newcommand{\cc}[1]{\mbox{\smaller[0.5]\texttt{#1}}}

\fvset{fontsize=\scriptsize,xleftmargin=8pt,numbers=left,numbersep=5pt}

\def\Snospace~{\S{}}

\newif\ifdraft\drafttrue
\newif\ifnotes\notestrue
\ifdraft\else\notesfalse\fi

\newcommand{\KL}[1]{\textcolor{Green}{KL: #1}}

\input{glyphtounicode}
\newcolumntype{R}[1]{>{\raggedleft\let\newline\\\arraybackslash\hspace{0pt}}p{#1}}

\newcommand{\squishlist}{
\begin{itemize}[noitemsep,nolistsep]
  \setlength{\itemsep}{-0pt}
}
\newcommand{\squishend}{
  \end{itemize}
}

\usepackage{tikz}
\newcommand*\Circ[1]{%
\begin{tikzpicture}[baseline=(C.base)]
\node[draw,circle,inner sep=0.2pt](C) {#1};
\end{tikzpicture}}

\newcommand*\BCirc[1]{%
\begin{tikzpicture}[baseline=(C.base)]
\node[draw,circle,fill=black,inner sep=0.2pt](C) {\textcolor{white}{#1}};
\end{tikzpicture}
}

\usepackage{xstring}
\newcommand{\PP}[1]{
\vspace{2px}
\noindent{\bf \IfEndWith{#1}{.}{#1}{#1.}}
}

\usepackage{xstring}
\newcommand{\PPP}[1]{
\vspace{2px}
\indent{\it \IfEndWith{#1}{.}{#1}{#1.}}
}

\gdef\therev{dfe759e}
\gdef\thedate{2017-05-24 21:27:57 -0400}

\begin{document}


\title{
  \Large
  \bf{
  \sys: Compositing Security Mechanisms through Diversification \protect\\
  (with Appendix)
  }
}

\ifdefined\DRAFT
 \pagestyle{fancyplain}
 \lhead{Rev.~\therev}
 \rhead{\thedate}
 \cfoot{\thepage\ of \pageref{LastPage}}
\fi


\author{
 Meng Xu,\; 
 Kangjie Lu,\;
 Taesoo Kim,\;
 Wenke Lee\;
\\
 \it{Georgia Institute of Technology}
}

\date{}
\maketitle

\begin{abstract} 
  A number of security mechanisms have been proposed
	to harden programs written in unsafe languages, each of which
	mitigates a specific type of memory error.  Intuitively, enforcing
	multiple security mechanisms on a target program will improve its
	overall security.  However,
	this is not yet a viable approach in practice because the execution
	slowdown caused by various security mechanisms is often
	non-linearly accumulated, making the combined protection
	prohibitively expensive; further, most security mechanisms are
	designed for independent or isolated uses and thus are often in
	conflict with each other, making it impossible to fuse them in a
	straightforward way.

	In this paper, we present \sys, an N-version-based system that
	enables different and even conflicting security mechanisms to be
	combined to secure a program while at the same time reducing the
	execution slowdown. In particular, we propose an automated
	mechanism to distribute runtime security
	checks in multiple program variants in such a way that conflicts
	between security checks are inherently eliminated and execution
	slowdown is minimized with parallel execution.  We also present an
	N-version execution engine to seamlessly synchronize these
	variants so that all distributed security checks work together to
  guarantee the security of a target program.  
\end{abstract}

\section{Introduction}
\label{s:intro}

Memory errors in programs written in unsafe languages (e.g., C/C++) have 
been continuously exploited by attackers~\cite{cvedetails}.
To defeat such attacks, the security community has 
deployed many security mechanisms such as widely deployed 
W$\oplus$X, which prevents code injection attacks by making Writable
memory not eXecutable, and ASLR, which prevents attacks (e.g., code
reuse) by making the address of target code/data
unpredictable. However, recent attacks~\cite{Shacham:2004,
  shacham:rop} have shown that these mechanisms are
not difficult to bypass. As such, more advanced 
techniques have been proposed. For example,
SoftBound~\cite{softbound}, CETS~\cite{cets}, and
AddressSanitizer~\cite{asan} (ASan) provide a high memory safety
guarantee, CFI~\cite{origCfi} and CPI~\cite{cpi} effectively mitigate
control flow hijacking attacks, MemorySanitizer~\cite{msan} (MSan) can
mitigate information leaks caused by uninitialized read, and
UndefinedBehaviorSanitizer~\cite{ubsan} (UBSan) can detect the causes of
undefined behaviors (e.g., null pointer dereference).

However, despite the large number of software hardening
techniques proposed, few of them actually get adopted in practice. 
One reason is that the slowdown imposed by these mechanisms erases
the performance gains that come from low-level languages.
Another reason is that each 
proposed technique tends to fix only specific issues while leaving the
program vulnerable to other attacks. Comprehensive security
protection is often demanded by mission-critical services such as web
servers or cyber-physical systems in which a single unblocked attack
could lead to disastrous consequences (e.g.,
heartbleed~\cite{heartbleed}).

In order to achieve 
comprehensive program protection, an intuitive method is to combine 
several techniques and enforce them together in a target program.
Unfortunately, this is often not viable in practice
for two reasons:
1) Runtime slowdown increases unpredictably after fusing different techniques.
For instance, in an already highly optimized build~\cite{cets}, combining
Softbound and CETS yields a 110\% slowdown--almost the sum of 
each technique individually;
2) Implementation conflicts prevent direct combination because most
techniques are not designed with compatibility in mind.
For instance, MSan makes the lower protected area
inaccessible, while ASan reserves the lower memory as shadow
memory. 
Re-implementing these techniques for better compatibility 
requires significant engineering effort if it is even possible.

In the meantime, hardware is becoming cheaper and more powerful.  The
increasing number of CPU cores combined with larger cache and memory size
keeps boosting the level of parallelism, making it practical to improve
software security through a technique known as N-version
programming~\cite{nversion, nversion-multicore, nversion-update,
  nversion-browser}. As part of this trend, the N-version scheme is particularly
suitable to multi-core architectures because replicas can run on cores in
parallel.
%

An N-version system typically requires careful construction 
of N variants that are both functionally similar in
normal situations and behaviorally different when under attacks.
Hence, although each program version may be vulnerable to certain
types of attacks, the security of the whole system relies on
the notion that an attacker has to simultaneously succeed in attacking
all variants in order to compromise the whole system.
This property of the N-version system gives us insight on how to
provide strong security to a program and yet not significantly
degrade its end-to-end performance.
That is, by distributing the intended security to N program variants
and synchronizing their execution in parallel, we can achieve the same
level of security with only a portion of its running time plus an overhead
for synchronization. Hence, the challenges lie in how to produce the program
variants in a principled way and how to synchronize and monitor their
executions efficiently and correctly.

In this paper, we introduce \sys, an N-version-based approach to
both minimize the slowdown caused by security mechanisms and
seamlessly unify multiple security mechanisms without re-engineering 
efforts to any individual mechanism.
In short, \sys splits the checks required by security mechanisms and
distributes them to multiple variants of a program in automated and principled 
ways to minimize execution slowdown. By synchronizing the execution
of these variants, \sys guarantees comprehensive security for the
target program.

While the N-version mechanism has been well studied primarily for
fault-tolerance~\cite{nversion, nversion-multicore, varan}, 
\sys aims at improving a program's
security and enabling the composition of multiple security mechanisms with 
automated protection distribution mechanisms. 
In addition, \sys is a practical system as it does not require
extra modification to the system or compilation toolchain.
\sys supports
state-of-the-art mechanisms like ASan, MSan, UBSan, Softbound, and CETS. We
have tested it on a number of C/C++ programs, including SPEC2006, SPLASH-2x,
PARSEC benchmarks, Nginx, and Lighttpd web servers. Through three case
studies, we show that 
1) the slowdown for ASan can be reduced from 107\% to 65.6\% and 47.1\% 
by distributing the sanity checks to two and three variants, respectively;
2) the slowdown for UBSan can be reduced from 228\% to 129.5\% and 94.5\%
by distributing the sub-sanitizers to two and three variants, respectively; 
and 
3) the time overhead for unifying ASan, MSan, and UBSan
with \sys is only 4.99\% more than the highest overhead of enforcing any of 
the three sanitizers alone.
In summary, our work makes the following contributions:
\squishlist
\item We propose an N-version approach to enable different 
	or even conflicting protection techniques to be fused for 
	comprehensive security with efficiency. 
\item We present an improved NXE design in terms of
  syscall hooking, multithreading support, and execution optimization.
\item We have implemented \sys and validated the effectiveness of \sys's
  NXE and the protection distribution mechanisms in
  amortizing the slowdown caused by state-of-the-art security
  mechanisms.
\squishend

The rest of the paper provides background information and compares 
\sys with related works (\autoref{s:background}), 
describes the design and implementation of \sys
(\autoref{s:design}, \autoref{s:impl}), presents evaluation
results (\autoref{s:eval}), discusses its limitations and improvements
(\autoref{s:discussion}), and concludes (\autoref{s:conclusion}).

\section{Background \& Related Work}
\label{s:background}

\subsection{Memory Errors vs. Sanity Checks}
\label{ss:mem-defense}

Memory errors occur when the memory is accessed in an unintended 
manner~\cite{Szekeres:2013}.
The number of reported memory errors is still increasing~\cite{cvedetails}
and severe attacks (e.g., heartbleed~\cite{heartbleed}) exploiting 
memory errors emerge from time to time.
We provide a taxonomy in~\autoref{tbl:taxonomy} to summarize the errors.
In particular, any vulnerability that may change a pointer unintentionally
can be used to achieve out-of-bound reads/writes. 
Use-after-free and uninitialized read are usually
caused by logic bugs (e.g., double-free and use-before-initialization) or 
compiler features (e.g., padding for alignment). 
Undefined behaviors can be triggered by various software bugs,
such as divide-by-zero and null-pointer dereferences.

\begin{table}
  \centering
  \footnotesize
  \begin{tabular}{l lll}
\toprule
  \textbf{Memory Error} &\textbf{Main Causes} & \textbf{Defenses}\\
  \midrule
  Out-of-bound r/w		& lack of length check & SoftBound~\cite{softbound} \\
  		 				& format string bug	& ASan~\cite{asan} \\
         & integer overflow & \\
						& bad type casting & \\
  \midrule
  Use-after-free		& dangling pointer & CETS~\cite{cets} \\
  						& double free & ASan~\cite{asan} \\
  \midrule
  Uninitialized read	& lack of initialization & MSan~\cite{msan} \\
						& data structure alignment & \\
						& subword copying & \\
  \midrule
  Undef behavior	& pointer misalignment & UBSan~\cite{ubsan} \\
						& divide-by-zero  & \\
						& null pointer dereference  & \\
\bottomrule                                     
\end{tabular}

  \caption{A taxonomy of memory errors. We assume the program is not malware.
  This taxonomy is mainly derived from two systematic survey 
  papers~\cite{Szekeres:2013, vanderVeen:2012}.
  }
  \label{tbl:taxonomy}
\end{table}

How to defend a program against memory errors has been extensively
studied in recent years.
For each category in~\autoref{tbl:taxonomy}, we are able to find corresponding
defenses.
In this paper, we are particularly interested in the
\textit{sanitizer}-style techniques because they thoroughly enforce sanity 
checks in the program to immediately catch memory errors before they are
exploited.

\subsection{N-version System}
\label{ss:nvariant}

The concept of the N-version system was initially introduced as a software 
fault-tolerance technique~\cite{nversion} and was later applied in enhancing 
software security~\cite{nversion-multicore, nversion-schedule, nversion-update,
nversion-browser}. 
In general, the benefit of the N-version system is that an attacker is required
to simultaneously compromise all variants with the same input in order to
take down the whole system.
To achieve this benefit, an N-version system should have at least 
two components: 
1) a variant producer that generates diversified variants based on 
pre-defined principles and 
2) an execution engine (NXE) 
that synchronizes and monitors the execution of all program variants.
We differentiate \sys with related works along these two lines of 
work.

\PP{Diversification}
Diversification techniques represent the intended protection goal of an
N-version system, for example, 
using complementary scheduling algorithms to survive concurrency 
errors~\cite{nversion-schedule}; 
using dynamic control-flow diversity and noise injection to thwart 
cache side-channel attacks~\cite{nversion-cache};
randomizing sensitive data to mitigate privacy 
leaks~\cite{nversion-data, shadowexec};
running multiple versions to survive update bugs~\cite{nversion-update};
using different browser implementations to survive vendor-specific 
attacks~\cite{nversion-browser}.
Diversification can also be done in load/run time such as
running program variants in disjoint memory layouts to mitigate code reuse 
attacks~\cite{nvariant, replicae, nversion-gadget}.

Differently, \sys aims to reduce the execution
slowdown and conflicts of security mechanisms. To achieve this goal,
two protection distribution mechanisms are proposed.
Partial DA checking~\cite{woda} also attempted to improve the performance of 
dynamic analysis with the N-version approach. However, it does not provide 
any protection distribution mechanism; instead it just insecurely skips 
checking some syscalls to improve performance. 
In addition, attacks can be completed by exploiting the vanilla variant 
before other protected variants find out.
In contrast, \sys proposes two principled diversification techniques 
to achieve this goal and also presents a more 
robust NXE with thorough evaluation.

\PP{NXE} Depending on how the program variants are generated, an NXE is 
designed to synchronize variants at different levels, such as
instruction/function level~\cite{nversion-schedule},
syscall level~\cite{nvariant, replicae, orchestra, varan, 
remon, mvarmor} or, file/socket IO level~\cite{nversion-browser}.
%
\sys shares some common features with other syscall-based NXE systems, 
including syscall divergence comparison (both sequence and arguments),
virtual syscall and signal handling, and a leader-follower execution pattern 
backed by a ring-buffer-based data structure for efficient event 
streaming~\cite{varan, remon, mvarmor}. 
However, \sys differs from these works in the following ways:

\PPP{Syscall hooking}
Prior works hook syscalls 
using a customized kernel~\cite{nvariant}, which jeopardizes its deployability; 
using the Linux \cc{ptrace} mechanism, which causes high synchronization 
overhead due to multiple context switches per each 
syscall~\cite{nvariant, replicae, orchestra, remon};
or binary-rewriting the program to redirect a syscall to a
trampoline~\cite{varan}, which may break the semantics of the program when 
replacing an instruction with fewer bytes with one with more bytes. 
%
MvArmor~\cite{mvarmor} leverages \cc{Dune}~\cite{dune} and Intel VT-x. 
However, it incurs a high overhead for syscalls that needs 
passthrough and is also subject to the limitations of \cc{Dune} (e.g., 
signal and threading).
To tackle these issues, \sys hooks syscalls by temporarily patching the 
syscall table with a loadable kernel module.

\PPP{Multithreading support}
Multithreading support varies in proposed NXEs, such as 
allowing only processes to be forked, not threads~\cite{nvariant};
enforcing syscall-ordering across the threads~\cite{orchestra, varan, mvarmor},
which can easily cause deviations, as not all threading primitives involves 
syscalls (e.g., \cc{pthread_mutex_lock});
and using CPU page fault exception to synchronize all memory 
operations~\cite{replicae}, which leads to high overhead.
None of these works are evaluated on multithreading benchmarks
like PARSEC or SPLASH-2x.
ReMon~\cite{remon} seems to have a similar level of support for multithreading 
as \sys---race-free programs by injecting synchronization agents into the
compiled binary, as discussed in~\cite{remon-details}.
\sys borrows the \textit{weak determinism} concept from the deterministic 
multithreading (DMT) domain and fully describes its design and implementation 
details to support it. \sys also identifies its limitations and potential 
solutions.

\subsection{Security/performance Trade-offs}
\label{ss:bg-trade-off}
Another approach to fit security into the performance budget is 
to devise a subset of protections from a full-fledged technique. 
Compared with Softbound~\cite{softbound}, which handles both code and data 
pointers, CPI~\cite{cpi} only instruments code pointers, 
which are less prevalent but more critical to 
code-reuse attacks~\cite{shacham:rop}.
Since the number of sanity checks inserted is dramatically reduced, 
CPI reduces the performance overhead from about 70\% to 8.4\%.
Similarly, ASAP~\cite{asap} keeps less commonly executed (i.e., less costly) 
checks and removes hot checks. 
However, selective protections sacrifice security. The assumption that security
is proportional to sanity check coverage is not valid in many cases.
More specifically, say a program contains two exploitable 
buffer overflow vulnerabilities; eliminating only one does not actually
improve the security, as one bug is enough for the adversary to launch the
attack. Unlike CPI and ASAP, \sys is a novel concept to reduce the slowdown 
caused by security mechanisms without sacrificing any security. 

\section{Design}
\label{s:design}

In a typical N-version system, program variants are executed in parallel and
synchronized by an execution engine to detect any behavior divergence.
The whole system terminates only when all variants have terminated. Therefore,
the overall runtime of an N-version system can be decomposed into two parts:
1) the time required to execute the slowest variant, 
and 2) any additional time used for variant synchronization and monitoring.

\subsection{Protection Distribution Principle}
\label{ss:design-model}

\sys's protection distribution applies to \textit{sanitizer}-style techniques,
which have three properties:
1) They enforce security via instrumenting the program with runtime
checks;
2) All the checks instrumented are independent of each other in terms of 
correctness; and
3) A sanity check alters control flow when and only when the check fails,
i.e., the program behaves normally when no memory errors or attacks are 
present.
The majority of memory error prevention techniques are
\textit{sanitizer}-style, including stack cookies, CFI, CPI, SAFECode, ASan, 
MSan, UBSan, Softbound, and CETS, which is also the foundation of 
profiling-guided security retrofitting such as ASAP~\cite{asap} and 
Multicompiler~\cite{multicompiler}.

These properties allow \sys to 
1) measure runtime overhead imposed by the sanitizer as well as remove sanity 
checks; 
2) split the program to allow only portions of the program to be 
instrumented with sanity checks;
3) split the set of security techniques to allow only selected checks
to be enforced; and 
4) produce functionally similar program variants such that \sys can synchronize 
their executions and reason about behavior divergences.
\sys distributes checks with two principles. 

\Srcdist takes a single security technique (e.g., ASan) as it is
and distributes its runtime checks on a program over N program variants.
Specifically, \sys first splits the program into several
disjoint portions and then generates a set of variants, each with only
one portion of the program instrumented by the technique.  
Since only a fraction of the code is instrumented with security
checks, the slowdown for each variant is smaller
compared with a fully instrumented program. And given that all portions of the
target program are covered through the collection of the variants, the
security protection is the same as if the security mechanism is
applied to the whole program.

In the example of ASan, the splitting unit is every function in the program 
and for a 3-variant split, after \srcdist, each variant has \sfrac{1}{3} 
functions instrumented with sanity checks and the other 
\sfrac{2}{3} uninstrumented, while collectively, all functions are covered.

It is worth noting that instrumentations added by sanitizers fall into 
two categories: 
metadata maintenance (e.g., bound and alias information in the example of
ASan or SoftBound) and sanity checks. \sys does not remove instructions
related to metadata maintenance, as removing them will break the correctness 
of a sanitizer. 

\Secdist takes multiple security techniques and distributes them 
over N variants. Specifically, \sys first splits these security mechanisms 
into several disjoint groups whereby each group contains security
mechanisms that are collectively enforceable to the program, i.e.,
they do not conflict with each other.  Since only a subset of the
intended protections is enforced on each variant, the
slowdown for each variant is smaller compared with the case
when all protection techniques are enforced on the same program (if ever 
possible).  Another important benefit of \secdist is that by distributing
security mechanisms to multiple program variants, any conflicts between
them (e.g., ASan and MSan) can be avoided without re-engineering these
security mechanisms. Since all intended protections are enforced
through the collection of variants, the overall protection is the same as
if all mechanisms are applied to the whole program.

In the example of UBSan, the splitting unit is every sub-sanitizer in 
UBSan, such as \cc{integer-overflow} and \cc{divide-by-zero}. 
For a 3-variant split, after \srcdist, each variant has a disjoint
set of sub-sanitizers (i.e., \cc{integer-overflow} appears in only one variant),
while collectively, all sub-sanitizers are covered.

Due to space constraints, interested readers may find a formal mathematical 
modeling of \sys's protection distribution principles 
in~\autoref{ss:supp-model}.

\subsection{Automated Variant Generator}
\label{ss:design-frontend}

\begin{figure}
  \centering
  \footnotesize
  \includegraphics[width=0.9\columnwidth]{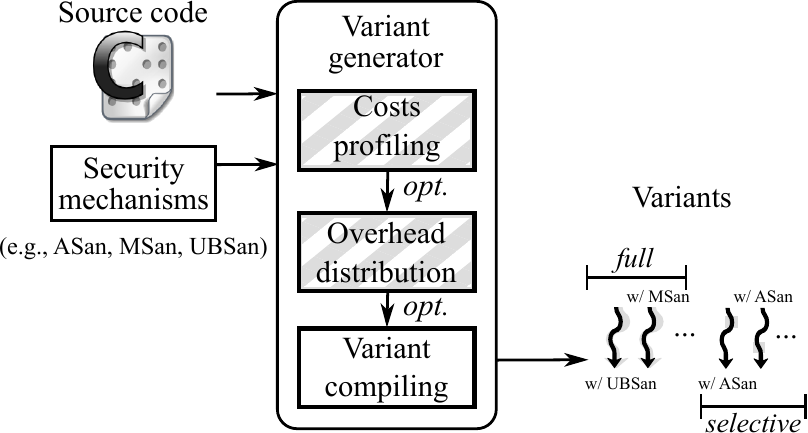}
  \caption{Variant generator workflow}
  \label{fig:variant-gen}
\end{figure}

\autoref{fig:variant-gen} illustrates the high-level workflow of the variant
generator.
The generator first compiles the target program without any security
mechanisms enforced and runs it with the profiling tool to get a baseline
profile.
In the \srcdist case, the generator then compiles and profiles the program with
the intended security mechanism.
In the \secdist case, the generator compiles and profiles the program multiple
times, each time with one of the intended security mechanisms.
The overhead profile is derived by comparing the security-enforced profiles
with the baseline profile.
In the next step, the generator runs the overhead distribution algorithm with
the intended number of splits (the N value) and creates N build configurations
for the compilers, each corresponding to one program variant.
The goal is to distribute the overhead measured by the profiling step
\emph{fairly} to each variants such that all variants finish execution
at approximately the same time.
Finally, the generator compiles the program N times, each with a different
build configuration, to get N program variants.

\PP{Profiling}
\sys relies on profiling to obtain the runtime slowdown numbers as the inputs
to the overhead distribution algorithm.
We choose to explicitly rely on profiling because it is a reliable way to
obtain the actual cost of a particular sanity check without making assumptions
about the nature of the program or the sanitizer.
It also takes in the effect of not only extra CPU cycles required to run the
check, but also the side effects on cache-line usage, register pressure, or
memory allocations.
However, the profiling approach does require an adequate and representative
workload to simulate the usage patterns in a production environment. 
Fortunately, for many projects, such a workload is often available in a 
form of test suites, which can be directly used to build a profile. 
More sophisticated profiling
tools~\cite{Arnold:2001, Duesterwald:2000:SPH} are orthogonal to \sys and can
be leveraged to improve the overhead profiling if necessary.
After profiling, the sanity checks are distributed to N variants in a way 
that the sum of overhead in each variant is almost the same.

\PP{Variant compiling}
Variant compiling for \srcdist is essentially a "de-instrumentation" process
that involves deleting the instructions that are only used for sanity checks
instrumented by sanitizers.
In order to collect such instructions for deletion, \sys uses data and
control dependence information maintained during the compilation process and
performs \textit{backward slicing} to automatically collect sanity
check-related instructions and discard them.
Variant compiling for \secdist is trivial, as it can be done by simply
compiling the program with the compilation settings a user would normally
use for those sanitizers, as long as the sanitizers used to harden the program
are collectively enforceable.

\subsection{N-version Execution Engine}
\label{ss:design-backend}

\sys's NXE synchronizes the executions of N program variants and makes them
appear as a single instance to any external entity. We present and justify
various design efforts to improve \sys's NXE in efficiency and robustness.

\begin{figure}
  \centering
  \footnotesize
  \includegraphics[width=0.9\linewidth]{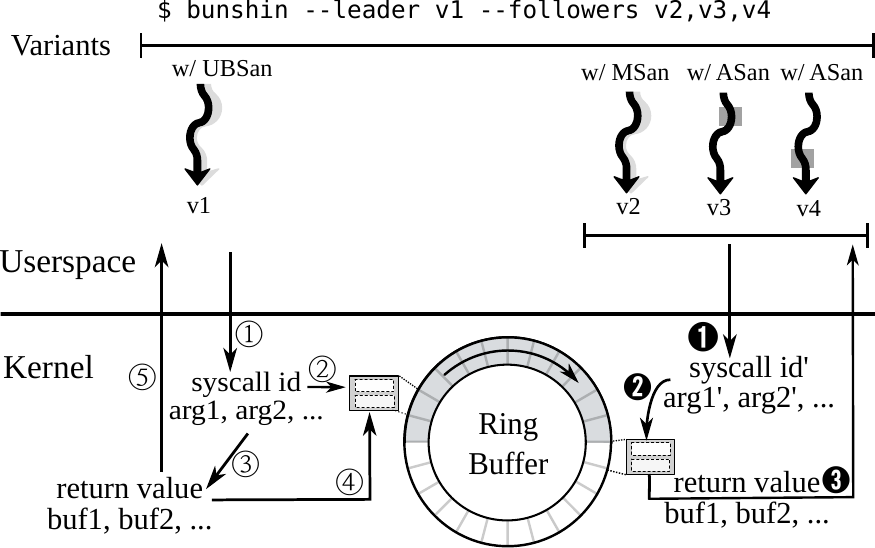}
  \caption{General synchronization procedure. The synchronization is triggered 
    when the syscall is trapped into kernel, as denoted 
    by \protect\Circ{1}\protect\BCirc{1} in both the leader and follower path.
  The leader then checks-in the syscall arguments to the shared 
  slot (\protect\Circ{2}), executes the syscall (\protect\Circ{3}), 
  and turns-in the 
  execution results in the shared slot (\protect\Circ{4}).
  A follower first checks whether the syscall arguments stored in the slot 
  match its own arguments (\protect\BCirc{2}) and if they match,
  directly fetches the results from sync slot (\protect\BCirc{3}) without
  actually performing the syscall. The difference between \textit{lockstep} 
  mode and \textit{ring-buffer} mode lies in whether step \protect\Circ{3} for 
  the leader can be performed before step \protect\BCirc{2} for all followers 
  is completed.}
  \label{fig:sync-modes}
\end{figure}

\PP{Strict- and selective- lockstep}
To synchronize the leader and the follower instances, a lockstep at
syscalls is required. \sys provides two lockstep modes:
\textit{strict-lockstep} and \textit{selective-lockstep}. 
In \textit{strict-lockstep} mode, the leader executes the syscall only if
all followers have arrived and agreed on the syscall sequence and arguments.
This ensures the security guarantee---the attack cannot complete in
either instance.
The downside is that variants are frequently scheduled in and out of the CPU due
to the necessary waiting, leading to higher runtime slowdown.

We observed that many attacks always trigger certain syscalls before
the actual damages are caused. For example,  with ASLR enabled,
attacks (e.g., ROP) generally leak an address first via I/O write
syscalls and then use the leaked address to construct subsequent attack payloads.
Based on this observation, \sys also provides the
\textit{selective-lockstep} so that users can choose to prevent the
attacks with higher performance. Specifically, \sys uses the 
\textit{ring-buffer} mechanism to synchronize instances---the leader 
executes at near full speed and
keeps dumping the syscall arguments and results into the shared ring buffer
without waiting, unless the buffer is full. The followers consume 
the syscall arguments and results at their own speed. Meanwhile,
lockstep is still enforced for the selected syscalls (e.g., write
related), as illustrated in~\autoref{fig:sync-modes}.
%
%
Our evaluation~\autoref{ss:eval-efficiency}, 
shows that \textit{selective-lockstep} reduces the synchronization overhead 
by 0.3\%-6.3\% compared with the \textit{strict-lockstep} mode.

In short, \textit{strict-lockstep} should perfectly preserve the security 
guarantee of the underlying sanitizer, while 
\textit{selective-lockstep} is an option we provide
when ASLR is enabled, as any remote code-reuse attacks (e.g., return-to-libc and
ROP) will have to first leak code/data pointers to bypass ASLR. 
\textit{Selective-lockstep} is able to stop such attacks by catching 
the leaks at I/O writes. 
In fact, any information leak attempt that involves a pointer will be 
detected at I/O writes.
A detailed analysis of the security guarantees 
provided by \sys is evaluated in~\autoref{ss:eval-security}.

\PP{Multi-threading}
\sys supports multi-process/thread programs by assigning each group of
leader-follower processes to the same \textit{execution group}, and each
execution group has its own shared buffers. The starting processes of
leader and follower variants form the first execution group, and when the
leader forks a child, the child automatically becomes the leader in the new
execution group. The child of a follower variant automatically becomes a 
follower in the new execution group.
In fact, for daemon-like programs, (e.g., Apache, Nginx, sshd),
simply separating parent and children processes into different
execution groups can be sufficient to eliminate syscall sequence
variations caused by non-deterministic schedulers because for those programs,
each thread/process is highly independent of the others and hardly ever or never
updates shared data.

However, for general-purpose multi-thread programs, synchronizing shared memory
accesses is necessary to ensure that the leader and followers have consistent
views on shared data. This can be achieved by enforcing 
all followers to follow exactly the same order of shared memory accesses as the
leader (\textit{strong determinism}), which can hardly be achieved without
a high performance penalty, as evidenced in the deterministic multi-threading
(DMT) domain~\cite{kendo, coredet}.
As a compromise, inspired by Kendo~\cite{kendo}, \sys ensures only
that all followers follow exactly the same order of all lock acquisitions as
the leader (\textit{weak determinism}). For example, if thread 1 in the leader
acquires a mutex before thread 2 passes a barrier, the same order
will be enforced in all followers. For programs without data
races, strong determinism and weak determinism offer equivalent 
guarantees~\cite{kendo}.
\sys achieves this with an additional 8.5\% overhead on SPLASH-2x and PARSEC
benchmarks (\autoref{ss:eval-efficiency}).

We argue that ensuring \textit{weak determinism} is sufficient for the majority of
multi-thread programs, as race-free programming is encouraged and tools have
been proposed to help developers eliminate data races~\cite{atomizer, IFRit}. 
However, should this becomes a problem in the future, \sys is capable of 
plugging in sophisticated DMT solutions such as DThreads~\cite{dthreads} with 
minor adjustments.

\PP{Sanitizer-introduced syscalls} Memory safety techniques generally issue
additional syscalls during program execution to facilitate sanity checks.
With all sanitizers we tested, i.e., ASan, MSan, UBSan, Softbound, CETS,
CPI, and SAFECode, all introduced syscalls can be categorized into three
classes:
1) pre-launch data collection,
2) in-execution memory management, and
3) post-exit report generation.
To illustrate, before executing the \cc{main} function, ASan goes through a
data collection phase by reading various files in \cc{/proc/self} directory
(on Linux system). During program execution, ASan issues
additional memory-related syscalls for metadata management. Upon program exit,
ASan might invoke external programs to generate human readable reports.

Given that variants instrumented with different sanity checks are
expected to have diverged syscall sequences, \sys needs to address
this issue to avoid false alerts.
In achieving this, \sys
1) starts synchronization only when a program enters its \cc{main} function;
2) ignores all the memory management-related syscalls; and
3) stops synchronization by registering as the first \cc{exit} handler.

We verified that all the syscall divergences caused by the
aforementioned sanitizers are successfully resolved. Although this is only
an empirical verification, we believe this can be a general solution
because any practical security mechanism should not alter program semantics
in normal execution states, which, reflected to the outside entities, are
syscall sequences and arguments.

\section{Implementation}
\label{s:impl}

\subsection{Automated Variant Generator}
\label{ss:impl-frontend}

\PP{Profiling}
To obtain overhead data for \srcdist, \sys instruments the program with
performance counters based on how the underlying sanitizer works.
As a prototype system, \sys currently measures the execution time of all program
functions based on the observation that the majority of memory-related security
checks (as discussed in~\autoref{ss:mem-defense}) operate at function level.
We discuss how to perform profiling instrumentation in a generic way
in~\autoref{s:discussion}.
Obtaining profiling data for \secdist is easy, as no extra instrumentation
is needed. \sys runs the program with each security mechanism individually
enforced and obtains the overall execution time.

\PP{Check removal}
\sys removes sanity checks at function level and the process 
consists of two steps:

In the \textit{discovery} step, \sys compiles a baseline version and
an instrumented version of the same program and then uses an analysis pass 
to dump the added/modified basic blocks per function. Among these basic blocks,
\sys considers a basic block that 1) is a branch target, 2) contains one of the 
known sanity check handler functions (e.g., in the case of ASan, functions 
prefixed with \cc{__asan_report_}), and 3) ends with the special 
LLVM \cc{unreachable} instruction as a \textit{sink} point for security 
checks. 
This is based on the properties of sanitizers, as a sanity check should 
preserve program semantics, i.e., special procedures are only invoked when 
a sanity check fails. Instrumentations for metadata maintenance involves 
neither sanity check functions nor \cc{unreachable} instructions and hence
are filtered out in this step.

In the \textit{removal} step, \sys automatically reconstructs sanity
checks based on the observation that sanity checks are instructions that
branch to the \textit{sink} points found in \textit{discovery} step.
After identifying the branching points and the corresponding condition
variables, \sys performs a recursive backward trace to variables and
instructions that lead to the derivation of the condition variable and
marks these instructions during tracing. The backward trace stops when
it encounters a variable that is not only used in deriving the value of the
condition variable but also used elsewhere in the program, an indication that
it does not belong to the sanity check. Removing the sanity check is achieved
by removing all marked instructions found in the above process.
This functionality is implemented as an LLVM pass.

\subsection{N-version Execution Engine}
\label{ss:impl-backend}

\PP{Pthreads locking primitives}
\sys enforces \textit{weak determinism} discussed 
in~\autoref{ss:design-backend} by re-implementing the full range of 
synchronization operations supported by \cc{pthreads} API, including locks, 
condition variables, and barriers.
\sys introduces a new syscall, \cc{synccall}, specifically for this purpose.
\cc{synccall} is exposed to processes under synchronization
by hooking an unimplemented syscall in an x86-64 Linux kernel (\cc{tuxcall}).
In the kernel module, \sys maintains an \cc{order_list} to record the
total ordering of locking primitive executions.
When a leader thread hits a primitive, e.g., \cc{pthread_mutex_lock},
it calls \cc{synccall} to atomically
put its \textit{execution group} id (EGID) in the \cc{order_list} and wake up
any follower threads waiting on its EGID before executing the primitive.
When a follower thread hits a primitive, the call to \cc{synccall} will first
check whether it is the thread's turn to proceed by comparing its EGID
and the next EGID in the \cc{order_list}. The thread will proceed if it matches;
otherwise, it puts itself into a variant-specific waitqueue. If the primitive
may cause the thread to sleep (such as a mutex), the thread wakes up its next
sibling in the waitqueue, if there are any, before sleeping.

Hooking \cc{pthreads}' locking primitives is done
by placing the patched primitives in a shared library, which is guaranteed to
be loaded earlier than \cc{libpthread}.

The drawbacks of this implementation are also obvious: 1) \sys is unable to
handle multi-thread programs that are built with other threading libraries or 
that use non-standard synchronization primitives (e.g., using \cc{futex} 
directly);
2) The performance overhead of \sys increases linearly with the usage
frequency of these primitive operations. Fortunately, the majority of 
multi-thread programs are compatible with \cc{pthreads} and locking 
primitives are only used to guard critical sections, which represent only 
a small fraction of execution.

\PP{Shared memory access}
Similar to the approach used by MSan to trace uninitialized memory accesses, 
whenever \sys detects mapping of shared memory into the variant's address 
space (by the indication of \cc{mmap} syscall with specific flag combinations), 
it creates shadow memory copies of the same size and then marks them as
"poisoned" state \cc{HWPOISON} whereby any access attempts to the mapped 
memory will also lead to an access attempt to the shadow copy, which 
eventually triggers a signal (\cc{SIGBUS}). Upon capturing the signal, 
memory access is synchronized in the normal way syscalls are handled, 
i.e., compare and copy content of the accessed address from the leader's 
mapping to the followers mapping.

\PP{Workflow}
\sys can be started with the path to each variant and the
program arguments. \sys first informs the kernel module to patch the syscall
table and then sets the \cc{LD_PRELOAD} environment variable to the library
containing patched virtual syscalls and pthread locking primitives.
It then forks N times and launches one program variant in each child process.
After that, \sys pauses and waits for status change of the variants. If 
any of the variant process is killed due to behavior divergences, \sys alerts
and aborts all variants. Otherwise, it exits when all variants terminate.

\section{Evaluation}
\label{s:eval}

In this section, we first evaluate \sys's NXE in terms of robustness and
efficiency. In particular, we run \sys on various programs and empirically
show that \sys is capable of handling the majority of them with low overhead
and no false alarms.
We also empirically test whether \sys can provide the same level of 
security guarantee as the underlying sanitizers, in other words,
whether \sys might compromise the security by partitioning the program
or splitting the sanitizers.

We then showcase how to accelerate ASan-hardened programs with \srcdist
and UBSan-hardened programs with \secdist.
We use another case study -- combining ASan, MSan, and UBSan --
to show that \sys is capable of unifying security mechanisms that
have conflicted implementations.

We also evaluate \sys in terms of hardware resource consumption, 
which could limit \sys's applicability, and report the performance of \sys 
under various levels of system load.

\PP{Experiment setup} The experiments are primarily conducted on 
a machine with Intel Xeon E5-1620 CPU (4 cores) and 64GB RAM running 64-bit
Ubuntu 14.04 LTS, except the experiment on scalability, which
is done with Intel Xeon E5-2658 (12 cores), and the experiment on RIPE
benchmark, which is done on a 32-bit virtual machine.
For evaluations on web servers, we dedicate another machine to launch
requests and measure server response time. The client machine is
connected to the experiment machine with a direct network cable. The
associated network card permits 1000Mb/s bandwidth.
Unless stated otherwise, the NXE is configured to run in 
\textit{strict-lockstep} mode for stronger security guarantee.

\subsection{NXE Robustness}
\label{ss:eval-sync}

We use a mixed sample of CPU-intensive and IO-intensive programs for
experiments, including SPEC2006 benchmark representing single-thread
programs, PARSEC, and SPLASH-2x benchmark for multi-threaded programs, and
Nginx and Lighttpd as representative server programs.
For each sample program, we compile it with the LLVM compiler framework and run
the same binary on \sys's NXE, i.e., \sys will synchronize identical N
binaries. This is to (empirically) verify the robustness of \sys's NXE design.

We do not observe false positives in any experiments on SPEC, SPLASH-2x, Nginx,
and Lighttpd. However, \sys is only able to run on six out of 13 programs
in the PARSEC benchmark. \cc{raytrace} would not build under \cc{clang} with
\cc{-flto} enabled. \cc{canneal}, \cc{facesim}, \cc{ferret}, and \cc{x264}
intentionally allow for data races. \cc{fluidanimate} uses ad-hoc
synchronization and hence, bypassing \cc{pthreads} APIs and \cc{freqmine}
does not use \cc{pthreads} for threading. These represent the limitation of
\sys's NXE: enforcing only weak determinism on \cc{pthreads} APIs.

\subsection{NXE Efficiency \& Scalability}
\label{ss:eval-efficiency}

\begin{figure}[!t]
  \vspace{-15pt}
  \centering
  \footnotesize
  \includegraphics[width=1.0\linewidth]{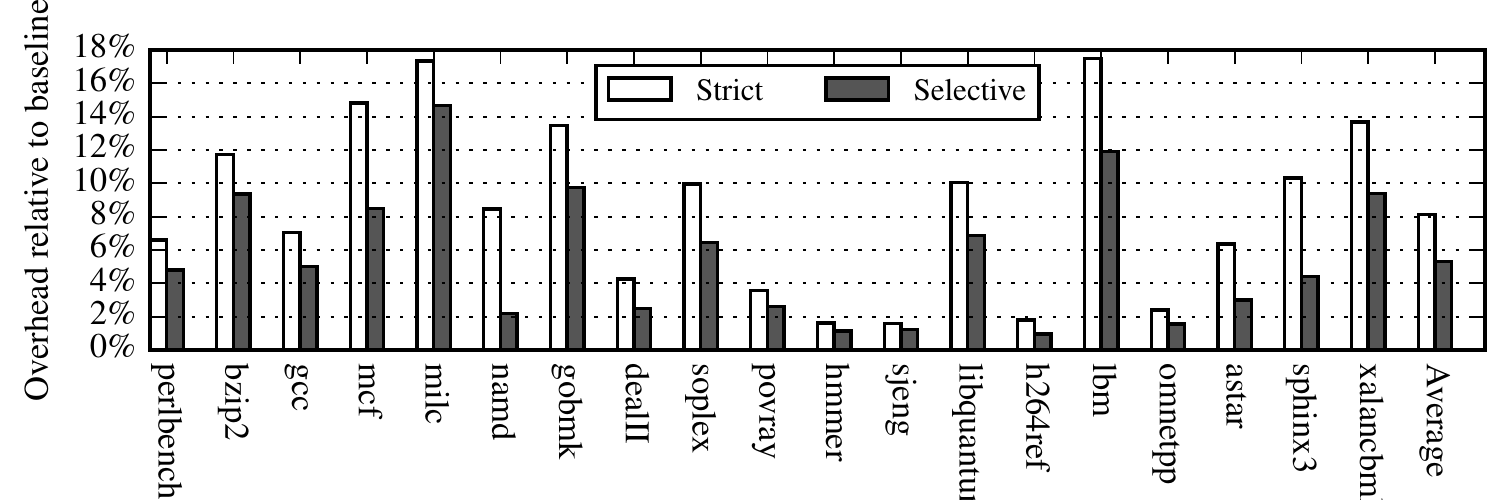}
  \caption{Evaluation of \sys' NXE efficiency with SPEC2006.}
  \label{f:spec-sync-eval}
\end{figure}

\begin{figure}[!t]
  \centering
  \footnotesize
  \includegraphics[width=1.0\linewidth]{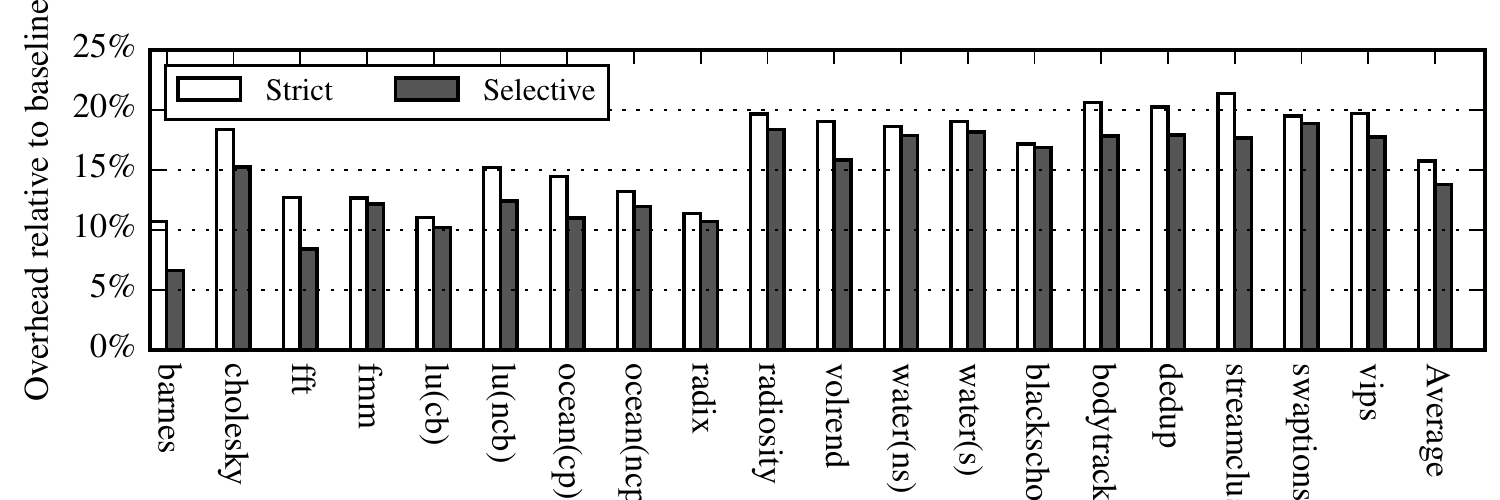}
  \caption{Evaluation of \sys's NXE efficiency with SPLASH-2x
  and PARSEC (number of threads = 4).}
  \label{f:spec-sync-multi}
\end{figure}

\autoref{f:spec-sync-eval} and \autoref{f:spec-sync-multi} show the efficiency 
evaluation of the NXE under both strict- and selective-lockstep modes when 
synchronizing 3 variants.
For the SPEC2006 benchmark, the average slowdowns for the two modes 
are 8.1\% and 5.3\%, respectively.
The overhead is sightly higher on multi-threaded benchmarks (SPLASH-2x and
PARSEC) -- 15.7\% and 13.8\%. This is due to the additional overhead for
recording and enforcing the total order of locking primitive acquisitions.
The selective-lockstep mode reduces the overhead by 0.3\%-6.3\% in the benchmark
programs.

We further evaluate the efficiency of \sys's NXE on
two server programs, \cc{lighttpd}, representing single-thread servers, and 
\cc{nginx}, representing multi-threaded servers.
We synchronize 3 variants and for \cc{nginx}, we run 4 worker threads, 
the default value after installation.
We simulate various workload situations by using 64 (light),
512 (heavy), and 1024 (saturated) concurrent connections and simulate
HTTP requests to files of 1KB and 1MB.

The results are shown in~\autoref{tbl:sync-server}. A noticeable difference
is that the percentage overhead when requesting small files (e.g., 1KB),
is significantly larger compared with requesting large files (e.g., 1MB).
The reason is that, while the absolute value of overhead is comparable in
both situations, it can be better amortized into the networking time of
a large file, therefore leading to smaller relative overhead.
We believe that in real-world settings when the servers are connected to LANs
and WANs, even the overhead for smaller files can be amortized in the
networking time, leading to unnoticeable overhead.

\begin{figure*}[!t]
  \vspace{-15pt}
  \centering
  \footnotesize
  \includegraphics[width=1.0\linewidth]{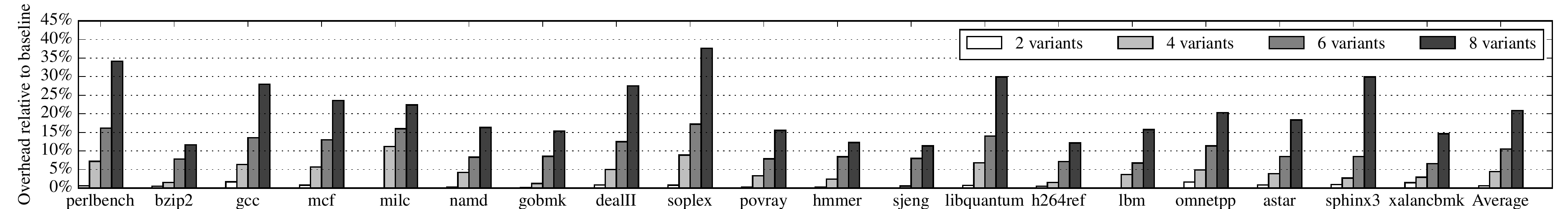}
  \caption{Scalability of \sys in terms of synchronizing 2 to 8 variants.
  For each program, we show the synchronization overhead over the 
baseline execution. On average, the overhead almost doubled with 2 more 
variants synchronized. Different programs show slightly different patterns in 
overhead growth. One of the reasons could be their differences in cache 
sensitivity~\cite{spec-cache}.}
  \label{f:spec-sync-scale}
\end{figure*}

\begin{table}[t]
  \vspace{-15pt}
  \centering
  \scriptsize
  \begin{tabular}{c | r | c | rr | rr}
	\toprule

  \textbf{Config} &
	\textbf{Conn} &
  \textbf{Base} &
  \multicolumn{2}{c}{\bf Strict}	& 
  \multicolumn{2}{c}{\bf Selective}	\\

  \midrule

  \cc{lighttpd} & 64    & 10.3  & 11.9  & 15.3\%  & 11.8  & 14.6\% \\
  1 Process   & 512   & 8.71  & 10.5  & 20.5\%  & 10.1  & 15.7\% \\
  1K File     & 1024  & 8.48  & 10.4  & 22.6\%  & 10.1  & 19.3\% \\

  \midrule

  \cc{lighttpd} & 64  & 974   & 994   & 2.05\%  & 992   & 1.85\% \\
  1 Process     & 512 & 959   & 972   & 1.35\%  & 970   & 1.15\% \\
  1M File      & 1024 & 955   & 964   & 0.91\%  & 961   & 0.63\% \\

  \midrule 

  \cc{nginx}    & 64  & 9.81  & 11.6  & 18.7\%  & 11.2  & 14.3\% \\
  4 Threads     & 512 & 8.46  & 10.3  & 21.9\%  & 9.88  & 16.8\% \\
  1K File      & 1024 & 8.20  & 10.2  & 24.4\%  & 9.63  & 17.4\% \\

  \midrule 

  \cc{nginx}    & 64  & 950   & 967   & 1.79\%  & 964   & 1.47\% \\
  4 Threads     & 512 & 985   & 999   & 1.40\%  & 996   & 1.12\% \\
  1M File     & 1024  & 979   & 998   & 1.94\%  & 995   & 1.63\% \\

  \midrule

  Ave. (1KB)    &     &       &       & 20.56\%  &       & 16.4\% \\
  Ave. (1MB)    &     &       &       & 1.57\%  &       & 1.31\% \\
  
  \bottomrule
  
\end{tabular}

  \caption{Performance of \cc{lighttpd} and \cc{nginx} under \sys's NXE,
  Performance measured as the processing time per request (unit. $\mu$s).
We use \cc{apachebench} as test driver and run each experiment 1000 times to
reduce the effect of network noise.}
  \label{tbl:sync-server}
\end{table}

\autoref{f:spec-sync-scale} shows the scalability of \sys's NXE in terms of 
total number of variants synchronized. We use a 12-core machine for this 
experiment as the number of variants should not exceed the number of cores
available.
As the number of variants goes from 2 to 8, the 
overhead increases from 0.9\% to 21\% accordingly. 
The primary reason for overhead increase is the LLC cache pressure, as
the cache miss rates increase exponentially when more variants are executed
in parallel. Recently added CPU features such as Intel Cache Allocation 
Technology~\cite{intel-cat} might help to mitigate this problem.

\subsection{Security Guarantee}
\label{ss:eval-security}
\sys does not remove any sanity checks, but only distributes them into 
multiple variants.
In \textit{strict-lockstep} mode, \sys should not compromise 
the intended security guarantee, as no variant can proceed with a syscall
without the arrival of other variants. 
Conceptually, the only way to compromise 
all variants is to launch an attack that is out of the protection scope of 
the underlying sanitizer.
Given that there is no attack window between variants, if an input 
sequentially compromises all variants without causing a divergence, 
it means the input bypasses all sanity checks. In this case, 
the attack will also succeed even if all checks are enforced in one variant 
(i.e., no \sys). In other words, the attack is essentially 
not in the protection scope of the underlying sanitizers and hence 
will not be in the scope of \sys.

On the other hand, the \textit{selective-lockstep} mode might introduce 
an attack window between the variants that allow 
an attacker to potentially compromise them one by one. However, 
\sys remains effective if the window is small enough.
To quantify the attack window, we measure the syscall distance between the 
leader and the slowest follower during our experiments. 
For CPU-intensive programs (SPEC2006, PARSEC, and SPLASH-2x), the average 
number of syscall gap is 5 while for IO-intensive programs 
(lighttpd and nginx), the average number of syscall gap is only 1.
The gap is small because even in \textit{selective-lockstep} mode, the 
variants are still strictly synchronized at IO-related syscalls.
We believe that this is a small enough time frame to thwart attackers.

To empirically confirm that real-world attacks can be thwarted even in 
\textit{selective-lockstep} mode, we first evaluated \sys on the RIPE 
benchmark with \srcdist on ASan. In particular, in compiling the programs 
generated by the RIPE benchmark, we go through the normal \srcdist procedure 
to produce two variants, and then launch and synchronize them with our NXE.
The results in~\autoref{tbl:ripe} confirm that \sys does not compromise
the intended security guarantee of ASan.

\begin{table}[t]
  \centering
  \scriptsize
  \begin{tabular}{c | r | r | r | r}
	\toprule

  \textbf{Config} &
	\textbf{Succeed} &
  \textbf{Probabilistic} &
  \textbf{Failed} &
  \textbf{Not possible} \\

  \midrule

  Default & 114 & 16 & 720 & 2990 \\
  ASan & 8 & 0 & 842 & 2990 \\
  \sys & 8 & 0 & 842 & 2990 \\

  \bottomrule
  
\end{tabular}

  \caption{We first run the RIPE benchmark on vanilla 32-bit Ubuntu 14.04 OS,
  and 114 exploits always succeed and 16 succeed probabilistically. After
adding ASan in the compilation, only 8 exploits succeed. After applying
\srcdist on the programs, still the same 8 exploits succeed.}
  \label{tbl:ripe}
\end{table}

\begin{table}[t]
  \centering
  \scriptsize
  \begin{tabular}{c | c | c | c | c}
	\toprule

  \textbf{Program} &
	\textbf{CVE} &
  \textbf{Exploits} &
  \textbf{Sanitizer} &
  \textbf{Detect} \\ 

  \midrule

  nginx-1.4.0 & 2013-2028 & blind ROP & ASan & Yes \\
  cpython-2.7.10 & 2016-5636 & int. overflow & ASan & Yes \\
  php-5.6.6 & 2015-4602 & type confusion & ASan & Yes \\
  openssl-1.0.1a & 2014-0160 & heartbleed & ASan & Yes \\
  httpd-2.4.10 & 2014-3581 & null deref. & UBSan & Yes \\

  \bottomrule
  
\end{tabular}

  \caption{Empirically test \sys's security guarantee with real-world programs
  and CVEs.} 
  \label{tbl:vulns}
\end{table}

To further verify this, we applied \sys to five real-world programs,
\cc{nginx}, \cc{cpython}, \cc{php}, \cc{openssl}, and \cc{httpd}, which 
contain known vulnerabilities that can be detected by ASan 
(to evaluate \srcdist) and UBSan (to evaluate \secdist). 
Similar to the RIPE benchmark case, we apply \sys on these 
vulnerable programs to produce two variants and later launch and synchronize 
them with our NXE. We then use the same exploit that triggered the warnings 
from ASan or UBSan to drive the program under \sys and check to see 
whether the same warnings are raised.
The result show that all exploitation attempts are detected 
(\autoref{tbl:vulns}).

For a concrete example, we applied \sys to \cc{nginx-1.4.0}, which 
contains bug CVE-2013-2028 that can be 
detected by ASan. We use \srcdist to produce two variants, A and B and 
use three published exploits \cite{brop, nginx-vul1, nginx-vul2} 
to test whether they can succeed in exploiting the 
vulnerable \cc{nginx} protected under \sys. The result shows that, when 
the overflow is triggered, variant A issues a \cc{write} syscall 
(trying to write to \cc{stderr}) due to ASan's reporting, 
while B does not. A further investigation on the protection distribution 
report shows that the vulnerable function \cc{ngx_http_parse_chunked} is 
instructed to be protected by variant A, which explains why variant A 
issues the \cc{write} syscall.

\noindent{\bf Attacking \sys}.
Given the attack window in \textit{selective-lockstep} mode, an attacker might 
be able to complete some simple attacks before detection, such as killing child 
threads/processes, closing file/sockets, or exhausting resources by allocating 
large chunks of memory, etc., provided that the attacker can inject shellcode 
or reuse program code to invoke the call and place the arguments of syscall 
in correct addresses.
An attacker might also launch denial-of-service attacks by sending compromised
variants into infinite loops that do not involve synchronize syscalls (in both
modes) or sleep/pause indefinitely (in \textit{selective-lockstep} mode).

Another attack vector is \sys's variant monitor.
For example, an attacker might intentionally crash \sys with unhandled 
non-deterministic sources such as uninitialized data (e.g., some encryption 
libraries intentionally use uninitialized data as a source of entropy, 
although such a practice is discouraged).
In addition, although we take care to keep the variant monitor simple and 
secure, it is not guaranteed to be bug-free. Therefore, if
an attacker compromises the variant monitor, he/she might be able to 
circumvent syscall synchronization.

\begin{figure*}[!t]
  \vspace{-15pt}
  \centering
  \footnotesize
  \includegraphics[width=1.0\linewidth]{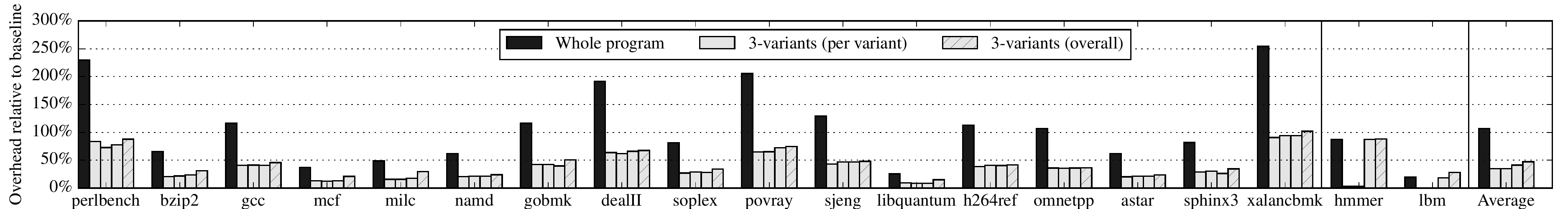}
  \caption{Effectiveness of \srcdist on ASan with three variants.
    For each program, we show the
  total overhead if ASan is applied to the whole program as well as
  per-variant overhead and \sys overall overhead. The two programs on the right are
  outliers that do not show overhead distribution.}
  \label{f:spec-split-asan}
\end{figure*}

\begin{figure*}
  \centering
  \footnotesize
  \includegraphics[width=1.0\linewidth]{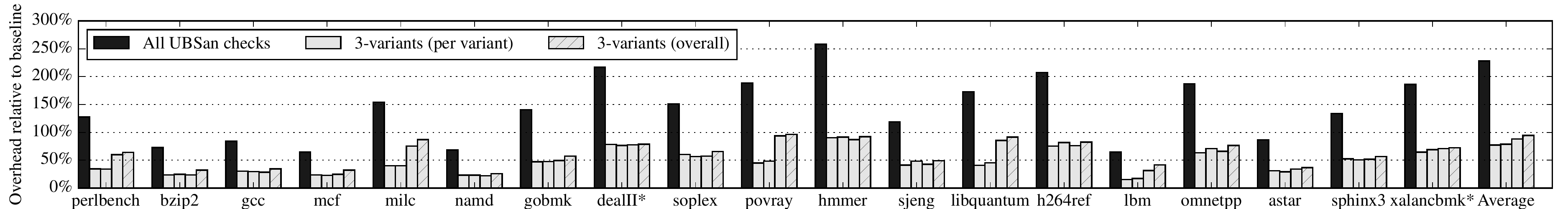}
  \caption{Effectiveness of \secdist on UBSan with three variants.
    For each program, we show the
  total overhead if all checks of UBSan are enforced as well as per-variant
overhead and \sys overall overhead. For \cc{dealII} and \cc{xalancbmk},
the overhead number is 4x larger than what is shown in the figure.}
  \label{f:spec-group-ubsan}
\end{figure*}

\subsection{\SrcDist on ASan}
\label{ss:eval-srcdist}

We show the effectiveness of \srcdist in accelerating the
performance of programs instrumented with ASan.
The reason we choose ASan for the case study is twofold:
1) ASan is representative of how memory error detection techniques are
generally enforced -- introducing runtime sanity checks.
In addition, the majority of checks placed in the program are independent of
each other and hence satisfy the assumption of \srcdist.
2) ASan provides a relatively high coverage on memory safety and hence is
appealing for long-living processes (like server programs) to thwart attackers
at runtime. However, the slowdown by enforcing ASan to the whole program is the
main obstacle in making it useful in production. We hope this
experiment will provide insights on how to use ASan through \sys.

The case study is done with the SPEC2006 benchmark programs using the \cc{train}
dataset for profiling and \cc{reference} dataset for the actual performance
measurement.
On average, the runtime slowdown caused by ASan is reduced from 107\%
(enforced to the whole program) to 65.6\% (2 variants) and
47.1\% (3 variants), respectively, about 11\% more than \sfrac{1}{2} and \sfrac{1}{3} of
the original slowdown. Due to space constraints, we show only results
for the more complex case (3 variants) in~\autoref{f:spec-split-asan}.

However, we also observed two outliers that do
not show overhead distribution: \cc{hmmer} and \cc{lbm}. After investigating
their execution profile, we observe that there is a single function that
accounts for over 95\% of the execution time and the slowdown caused by ASan.
Since \sys performs sanity check distribution at the function level, the overhead is
inevitably distributed to one variant, causing that variant to be the
bottleneck of the entire system. However, concentrating functionalities in one
single function is rarely seen in the real-world software we tested, including 
Python, Perl, PHP Apache httpd, OrzHttpd, and OpenSSL; hence, we do not 
believe these outliers impair the practicality of \sys.

\subsection{\SecDist on UBSan}
\label{ss:eval-secdist}

UBSan is a representative example to illustrate why collectively enforcing
lightweight sanity checks might lead to significant overall slowdown.
UBSan contains 19 sub-sanitizers, each with overhead no more than 40\%.
However, adding them leads to over 228\% overhead on SPEC2006 benchmarks,
making UBSan a perfect example to exercise \secdist.

Similar to the ASan case study, the test case is done with SPEC2006 programs
using \cc{train} dataset for profiling and \cc{reference} dataset for
experimentation.
On average, the runtime caused by UBSan is reduced from 228\%
(enforced all checks) to 129\% (2 variants) and 94.5\%
(3 variants), respectively, about 15\% more than \sfrac{1}{2} and \sfrac{1}{3} of
the original slowdown. Due to space constraints, we show only the results
for the more complex case (3 variants) in~\autoref{f:spec-group-ubsan}.
This deviation from the theoretical optimum is a bit larger compared with the ASan
case study because we only have 19 elements in the set and
hence are less likely to get balanced partitions across variants. 
However, it still shows the effectiveness
of \secdist in accelerating the overall performance.

\begin{figure*}
  \vspace{-15pt}
  \centering
  \footnotesize
  \includegraphics[width=1.0\linewidth]{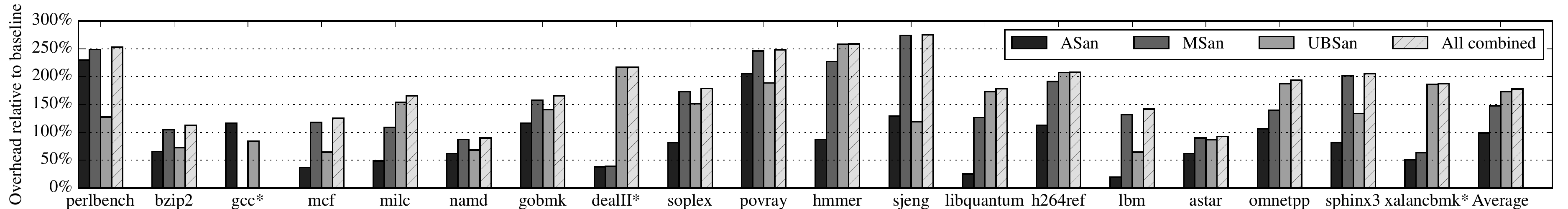}
  \caption{Performance result of each LLVM sanitizer respectively as well as
  the overall performance overhead when unified under \sys.
  \cc{gcc} cannot run with MSan, therefore, we exclude the evaluation on it.
For \cc{dealII} and \cc{xalancbmk} the overhead number is 4x larger than what
is shown in the figure.}
  \label{f:spec-conflict-llvm}
\end{figure*}

\begin{figure*}[!t]
  \centering
  \footnotesize
  \includegraphics[width=1.0\linewidth]{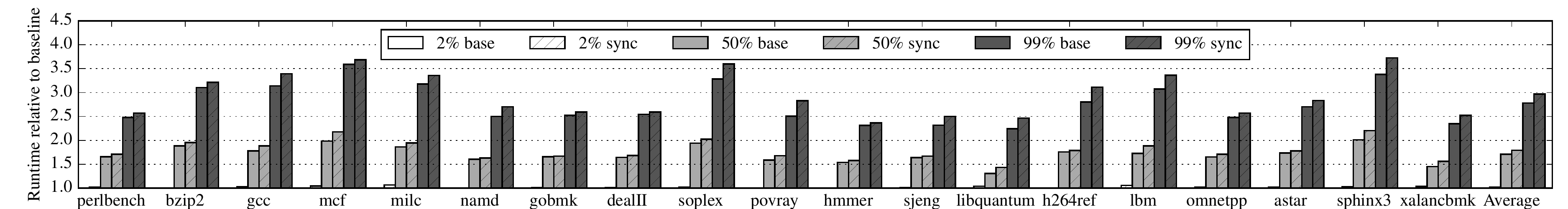}
  \caption{Evaluation of \sys execution engine under various workload levels.
  The experiment is done with the configuration of synchronizing 2 variants.
We use the system-stressing tool \cc{stress-ng} to add background
workloads including CPU tasks, cache thrashing, and memory allocations and
deallocations. We maintain the background load level at 50\% and 99\%,
respectively. The 2\% load for the baseline case is due to the kernel and OS
background services.}
  \label{f:spec-load-test}
\end{figure*}

\subsection{Unifying LLVM Sanitizers}
\label{ss:eval-union}

In theory, \sys is capable of unifying any security mechanism that falls in the
\textit{sanitizers} definition in~\autoref{ss:design-model}.
The reason we choose LLVM sanitizers (ASan, MSan, UBSan) for the case
study is mainly because: collectively, they provide almost full protection
against memory error, which we have not seen in any other work. Unifying them
through \sys might give some insight on how to achieve full memory error
protection without any re-engineering effort to these sanitizers. 

In this case study, each variant is simply the program compiled with one of the
sanitizers with the default compilation settings. We measure the execution time
of each program variant when running by itself and also the total execution
time of \sys. The result is reported in~\autoref{f:spec-conflict-llvm}.
On average, the total slowdown of combining these sanitizers is 278\%,
with only 4.99\% more compared with merely enforcing the slowest sanitizer
among the three. In other words, paying a little slowdown for synchronization
helps bring additional protection provided by the other two sanitizers.

\subsection{Hardware Resource Consumption}
\label{ss:eval-other-hw}

\PP{Memory}
Since all variants are loaded into memory for parallel execution, the basic
memory usage is almost linear to the number of variants. This is an inherent
trade-off for execution time. In addition, whether \srcdist helps
to split memory overhead caused by a sanitizer depends on the sanitizer'
internal working, In the case study of ASan, although each variant
executes only a portion of the sanity checks, it still needs to shadow the whole
memory space as required by ASan. Therefore, the memory overhead of ASan still
applies to each variant. However, the memory overhead can be distributed for
shadow stack-based techniques. By definition, \secdist can be used to
distribute memory overhead to multiple variants. In the UBSan case study,
the memory overhead of each variant is the sum of all enforced sub-sanitizers'
overhead.

\PP{CPU cycles}
\sys's NXE utilizes spare cycles in a multi-core CPU for efficient
variant synchronization. If the CPU does not provide sufficient
parallelism, \sys will not be able to improve the performance; instead, it
will only introduce more performance overhead. An evaluation on C/C++ programs
in the SPEC2006 benchmark shows that the average synchronization overhead is
103.1\% when running \sys on a single core.

Although \sys is not suitable for devices with a single core,
it does not mean that \sys requires exclusive cores to work.
In fact, due to OS-level task scheduling, \sys can exploit free cycles in
the CPU as long as not all cores are fully utilized.
\autoref{f:spec-load-test} shows that \sys's performance is stable under
various load levels. The average slowdown due to synchronization is
10.23\% and 13.46\%, respectively, when the CPU is half and fully loaded, slightly
higher than the case when the load is small (8.1\%).
The results prove that the performance of \sys is stable
across various load levels.

\section{Discussion}
\label{s:discussion}

\PP{Trading-off resources for time}
%
There is no doubt that \sys's parallelism consumes more hardware
resources; hence \sys is not suitable for cases where hardware resources are
scarce. In fact, \sys's design is inspired by the popularity of multi-core
processors and large-size cache and memory, and trade-off resource usages
for execution time.
\sys empowers users to make use of available
hardware resources to improve both security and runtime performance and
sheds lights on how to solve a difficult problem
---speeding up hardened programs without sacrificing security
---with simply more hardware resources, which are easy to obtain.

\PP{Sanitizer integration}
\sys currently has no integration with the sanitizers, i.e., it does not 
require detailed knowledge of how a sanitizer works in order to 
"de-instrument" the sanity checks. Although this gives \sys great flexibility,
it also prevents \sys from further optimization. 
For example, ASan still shadows the whole memory space even when only 
a subset of sanity checks is performed per program variant, thus leading
to increased memory usage in every variant.
To solve this, we could modify ASan's logic in a way such that only a 
portion of the memory space is shadowed in each variant; 
in other words, we can distribute the memory overhead to all program 
variants as well.

\PP{Finer-grained sanity check distribution}
As shown in the case of \cc{hmmer} and \cc{lbm}, sanity check distribution at 
the function level might be too coarse grained if one or a few functions dominate 
the total execution. Therefore, to enable finer-grained overhead 
distribution, we plan to look into performing both profiling and check removal 
at the basic block level. 

\section{Conclusion}
\label{s:conclusion}
We presented \sys, an N-version system that seamlessly 
unifies different and even conflicting protection techniques while at same time 
reducing execution slowdown.
\sys achieves this with two automated variant generation strategies 
(\srcdist and \secdist) for distributing security
checks to variants and an efficient parallel execution engine that synchronizes
and monitors the behaviors of these variants.
Our experiment results show that \sys is a practical system that can 
significantly reduce slowdown of sanitizers (e.g., 107\% to
47.1\% for ASan, 228\% to 94.5\% for UBSan) and 
collectively enforce ASan, MSan, and UBSan without conflicts
with only 4.99\% incremental overhead.

\section{Acknowledgment}
\label{s:ack}

We thank our shepherd, Ittay Eyal, and the anonymous reviewers for their
helpful feedback.
This research was supported by 
NSF under award DGE-1500084, CNS-1563848, CRI-1629851, 
CNS-1017265, CNS-0831300, and CNS-1149051,
ONR under grant N000140911042 and N000141512162,
DHS under contract No. N66001-12-C-0133,
United States Air Force under contract No. FA8650-10-C-7025,
DARPA under contract No. DARPA FA8650-15-C-7556,
and DARPA HR0011-16-C-0059,
and ETRI MSIP/IITP[B0101-15-0644].

\footnotesize
\bibliographystyle{plain}
\setlength{\bibsep}{3pt}
\bibliography{p,sslab,conf}

\newpage
\appendix

\section{Formal Modeling}
\label{ss:supp-model}

In this section, we present a formal modeling of the protection distribution
principles of in an N-version system, from which we derive the design goals 
for the two components of \sys: an automated variant generator and an 
efficient NXE.

\abovedisplayskip=3pt
\abovedisplayshortskip=3pt
\belowdisplayskip=3pt
\belowdisplayshortskip=3pt

\subsection{N-version system}
Since \sys is based on N-version systems, we first give a mathematical 
abstraction of the N-version system regarding runtime slowdown.

In a typical N-version system, program variants are executed in parallel 
and synchronized by an execution engine to detect any behavior divergence. 
The whole system terminates only when all variants have terminated. 
Therefore, the overall runtime of an N-version system can be decomposed 
into two parts: 
1) the time required to execute the slowest variant; and
2) any additional time used for variant synchronization and monitoring.
Expressed in terms of overhead (i.e., slowdown),
\begin{equation} \label{eq:nversion}
O_{bunshin} = max(O_{V_1}, O_{V_2}, ..., O_{V_N}) + O_{sync}
\end{equation}
It is obvious that an efficient NXE will help reduce $O_{sync}$ and the overall
slowdown. We further derive how to reduce the first term with variant
generation.

\subsection{\Srcdist} 
We formally model the \srcdist concept as
follows: We use $S$ to denote a \textit{sanitizer}, $U_i$ represents
a unit of program to be instrumented under $S$ for sanity checks.
$O_i$ represents the amount of overhead added by the instrumentation.
There are $|U|$ program units to be instrumented under $S$ in total.
The slowdown if all program units are instrumented is represented as
$O_{total} = \sum_{i=1}^{|U|} O_i + O_{residual}$ where $O_{residual}$
represents the slowdown that cannot be amortized by partitioning the program.

In the example of ASan, $U_i$ can be a C/C++ function in the program,
$O_i$ represents the slowdown implied by merely instrumenting the function
with runtime checks and $O_{residual}$ represents the slowdown caused by
metadata creation, book-keeping and report generation, which cannot be
distributed to multiple program variants.

\subsection{\Secdist} 
We formally model the \secdist concept as follows:
We use $C=<S_1, S_2, ..., S_{|C|}>$ to denote a collection of
\textit{sanitizers} that a user wants to enforce on the program. $O_i$ represents
the amount of slowdown incurred by enforcing $S_i$. Empirically,
the slowdown if all security mechanisms are enforced on the same program:
$O_{total} = \sum_{i=1}^{|C|} O_i \pm O_{synergy}$ where $O_{synergy}$
represents the additional overhead (or gain) incurred when two or more
mechanisms are combined.

In the example of UBSan, $S_i$ represents each sub-sanitizer that
constitutes UBSan, such as \cc{integer-overflow}, and \cc{divide-by-zero}, etc. 
$O_i$ represents the slowdown introduced by
only enforcing $S_i$ on the program. $O_{synergy}$ in this case is negative
(i.e., gain), as we observe that the combined sanitizer shares the
metadata creation and reporting operations, and hence, is considered 
performance gain.

\subsection{Variant Generation}
Given that both $O_{residual}$ and $O_{synergy}$ are orthogonal to \sys design,
(because \sys has no control over their values), we define the general
variant generation strategy as follows:

We use $P_i$ to denote a general unit of protection: in the case of \srcdist,
$P_i = U_i$ and in the case of \secdist, $P_i = S_i$. Protection distribution
for both \srcdist and \secdist can be unified as following:
Given a set of security checks, $P = <P_1, P_2, ..., P_K>$ and an
integer $N$, partition the elements in $P$ into $N$ subsets $V_1, V_2, ..., V_N$
in a way that
\begin{equation} \label{eq:1}
\bigcup_{i=1}^{N} V_i = P 
\quad \textrm{and} \quad
V_i \cap V_j = \emptyset, \forall i, j \in {1,2, ..., N}, i \ne j
\end{equation}
i.e., all units of the program
(in the case of \srcdist) or all security mechanisms
(in the case of \secdist) are covered without overlap.

As the goal of \sys is to distribute protection based on slowdowns,
we represent each $P_i$ by its overhead number $O_{P_i}$ and each
variant $V_i$ by $O_{V_i}$. Then, by~\autoref{eq:1}, we have
\begin{equation} \label{eq:3}
\sum_{i=1}^{N} O_{V_i} = \sum_{P_i \in P} O_{P_i} = O_{total}
\end{equation}
To minimize the first term in~\autoref{eq:nversion}, we need to:
\begin{equation} \label{eq:4}
minimize \quad \sum_{i=1}^{N} |O_{V_i} - \frac{O_{total}}{N}|
\end{equation}
In plain words, \sys needs to fairly distribute the total slowdown to all
variants so that every variant runs at approximately the same speed.
This is essentially finding the optimal solution to the abovementioned 
set N-partition problem.
However, being an NP-complete problem~\cite{np-prob}, any algorithm that
derives an optimal solution will take exponential time in relation to the
number of sanity checks, making it impractical.
For example, in the case of \srcdist on ASan, a program is partitioned by
functions and production software like \cc{nginx} contains about
2,000 functions. Hence, to be practical, \sys adopts a fast set-partitioning
algorithm that produces a near-optimal solution with polynomial time
complexity~\cite{algo-book}.

\subsection{Conclusion}
As the goal of \sys is to reduce total slowdown in~\autoref{eq:nversion},
we have two quantifiable ways to achieve this goal:
1) fair distribution of sanity checks across variants - this becomes the design
goal for the variant generator, and
2) reducing synchronization and monitoring overhead - this becomes the design
goal for the execution engine.
Correspondingly, we can derive the evaluation method for the two components:
\begin{itemize}
\item For the variant generator, measure $O_{V_1}$, $O_{V_2}$, ..., $O_{V_N}$ and
compare their maxima to the theoretical optimal $O_{total} / N$.
\item For the execution engine, use the same program binary as
$V_1$, $V_2$, ..., $V_N$ which eliminates the first term by making
$O_{V_1}=O_{V_2}=...=O_{V_N}$ and hence, any measurable overhead becomes the
synchronization overhead, i.e., $O_{total}=O_{sync}$.
\end{itemize}

\end{document}